\documentclass[amsmath,amssymb, aps, prb, twocolumn, notitlepage]{revtex4-2}
\usepackage{graphicx}
\usepackage{dcolumn}
\usepackage{bm}
\usepackage[usenames]{color}
\usepackage{slashed}
\usepackage[utf8]{inputenc}
\usepackage{amsfonts}
\usepackage{amsmath}
\usepackage{amssymb}
\usepackage{hyperref}
\usepackage{latexsym}
\usepackage{color}
\usepackage{mhchem}
\usepackage{braket}
\usepackage{diagbox}
\usepackage{booktabs}
\usepackage{MnSymbol}
\usepackage{float}
\makeatletter
\let\newfloat\newfloat@ltx
\usepackage{algorithm}
\usepackage{algorithmicx}
\usepackage{algpseudocode}

\newcommand{\printfnsymbol}[1]{%
  \textsuperscript{\@fnsymbol{#1}}%
}

\begin{document}
\title{1D Spontaneous Symmetry Breaking in thermal equilibrium via Non-Hermitian Construction}

\author{Jia-Bao Wang}
\affiliation{International Center for Quantum Materials, School of Physics, Peking University, Beijing, 100871, China}
\author{Zi-Hao Dong}
\affiliation{International Center for Quantum Materials, School of Physics, Peking University, Beijing, 100871, China}
\author{Yi Zhang}
\email{frankzhangyi@gmail.com}
\affiliation{International Center for Quantum Materials, School of Physics, Peking University, Beijing, 100871, China}

\date{\today}

\begin{abstract}
Spontaneous symmetry breaking generally circumvents one-dimensional systems with local interactions in thermal equilibrium. Here, we analyze a category of one-dimensional Hermitian models via local non-Hermitian constructions. Notably, spontaneous symmetry breaking and long-range order may emerge at finite temperatures in such systems under periodic boundary conditions, in sharp contrast to Hermitian constructions. We demonstrate clear numerical evidence, such as order parameters and specific heat, supporting phase diagrams with robust ordered phases. Non-Hermitian physics plays a vital role in prohibiting domain-wall proliferation and promoting spontaneous symmetry breaking. The fermions exhibit an exotic topological nature in their path-integral windings, which uphold nonzero integers - commonly a non-Hermitian signature - in the ordered phases, thus offering a novel and spontaneous origin for both symmetry breaking and non-Hermiticity. 
\end{abstract}

\maketitle

\emph{Introduction}--- Spontaneous symmetry breaking (SSB) is a central concept in particle and condensed matter physics \cite{mottbook, Ashcroft76, Altland_Simons_2010, negelebook, mahan2000book, fradkin2013book, nagaosa2013book, parisi1998statistical}, where the low-energy physics displays lower symmetry than the parent Hamiltonian, e.g., a ferromagnet, despite the parent Hamiltonian being spin symmetric, or a superconductor with broken `gauge' symmetry, despite the parent Hamiltonian being manifestly gauge invariant. Due to proliferating defects and domain walls, SSB and long-range order do not materialize in one-dimensional (1D) systems with local interactions in thermal equilibrium \cite{mottbook, Ashcroft76, Altland_Simons_2010, parisi1998statistical, MerminWagner1966, Hohenberg1967, Cardy1996, Halperin2019}.  On the other hand, exceptions exist in classical systems far from thermal equilibrium, as SSB in 1D dynamical, diffusive and driven systems \cite{SSB1D1995, SSB1D1986, SSB1D1998, SSB1D1997, Non-reciprocal}. 

There has been much recent progress on non-Hermitian quantum systems, such as non-Hermitian topological phenomena \cite{gong2018, kawabataprx,shen2018,kunst2018bi,hu2021knot,hu2022knot,li2022braiding,guo2023knot,fu2024braiding,nakamura2024,verma2024,liu2024localized, yao2018,origin2020}, the non-Hermitian skin effect (NHSE) \cite{yao2018, yokomizo2019, yang2020, zhang2020, origin2020}, and nontrivial path-integral winding \cite{hu2023nontrivial, hu2024residue}. 
 Another prominent topic is the $\mathcal{P}\mathcal{T}$-symmetry breaking in non-Hermitian physics, characterized by a spectral transition from purely real eigenvalues to complex-conjugate pairs at critical parameter values \cite{bender1998real, guo2009observation, el2018non, chong2011pt}.
While quantum mechanics dictates Hermitian Hamiltonians, the non-Hermitian models are increasingly proving their relevance in the fields of condensed matter physics \cite{bergholtzrev2021, ashida2020}, finding their origins in open or dissipative systems \cite{open1, open2, open3, open4, open5, open6, open7, open8}, optical \cite{optical1, optical2, optical3, optical4, optical5, optical6}, acoustic \cite{ding2016,yves2017,ding2018,tang2020,puri2021,tang2021,tang2022,zhang2023exp}, electric \cite{circuit1, circuit2, circuit3, circuit4, circuit5, circuit6}, mechanical \cite{brandenbourger2019,ananya2020,chen2021,wang2022morphing,wang2023exp,li2024obser}, and solid-state systems \cite{WeiChen2024}. 
Recently, non-Hermitian physics of non-reciprocal classical fields has laid the foundations for understanding SSB in dynamical and non-equilibrium systems \cite{Non-reciprocal}.

Here, we re-visit SSB in 1D quantum systems from the perspective of non-Hermitian constructions. We study 1D quantum systems where non-Hermitian fermions are locally coupled to a discrete field. The overall quantum systems are fully Hermitian and efficient for numerical and sign-problem-free Monte Carlo (MC) calculations. Interestingly, despite their 1D nature and local interactions, such systems exhibit clear signatures of long-range order following SSB, such as order parameters and specific heat behaviors, as the temperature is lowered. Therefore, for the first time, we deliver a 1D SSB scenario, which, unlike previous examples, is under thermal equilibrium and beyond classical dynamics. 

The reasons that our constructions enable 1D SSB with local interactions and thermal equilibrium are as follows. From a energetic perspective, non-Hermitian systems exhibit unique non-locality - the NHSE shows that boundaries or domain walls impact their physics globally \cite{yao2018, yokomizo2019, yang2020, zhang2020, origin2020}, and non-Hermitian delocalization yields long-range correlations and entanglements \cite{hu2023nontrivial, hu2024residue} - therefore, non-Hermitian constructions may effectively introduce attractive strings between domain walls that contain their proliferation. From a fermionic perspective, the phase transition accompanies a characteristic and essentially topological shift of fermion path integrals from conventional ($|w|=0$) to nontrivial (simultaneous $|w|=2, 4, \cdots$) path-integral winding-number across the periodic boundary conditions (PBCs). Results from numerical calculations and mean-field theories also portray consistent physics. Although overall Hermitian, various non-Hermitian properties, e.g., such as nonzero winding number and nontrivial fermion spectral functions \cite{supp}, emerge in the ordered phase, exhibiting what we dub spontaneous  `Hermiticity' breaking - a novel route toward non-Hermitian physics.

\begin{figure}
    \centering
    \includegraphics[width=0.5\textwidth]{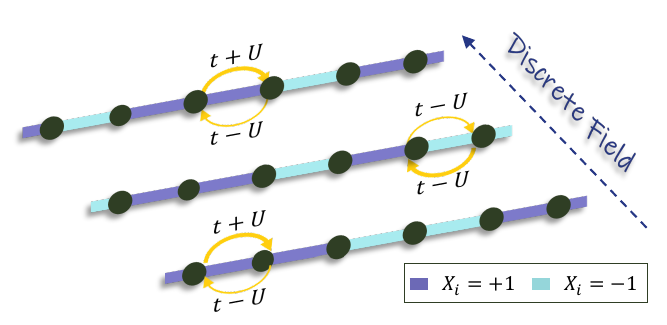}  
    \caption{The model in Eq. \ref{eq:ham} consists of fermions on the sites (black dots) and discrete $X_i=\pm 1$ (purple and cyan, respectively) on the bonds, which we sample or sum over for thermal equilibrium. The fermions' nearest-neighbor hopping is non-reciprocal: $t\pm U$ to the right and $t\mp U$ to the left for $X_i=\pm 1$. More generically, we include an Ising interaction $J$ between nearest-neighbor $X_i$ and next-nearest-neighbor hopping $t'$ of the fermions, which are not shown in the schematics. }
    \label{Fig:demonstration}
\end{figure}

\emph{Models}--- We consider the following 1D Hamiltonian under PBCs: 
\begin{eqnarray}
    \hat{H} &=& t\sum_{i} (c^\dagger_{i+1} c_i + c^\dagger_i c_{i+1}) + t'\sum_{i} (c^\dagger_{i+2} c_i + c^\dagger_i c_{i+2}) \nonumber \\
   &+& U\sum_{i} X_{i}(c^\dagger_{i+1} c_i - c^\dagger_i c_{i+1}) - J\sum_i X_i X_{i+1}, 
   \label{eq:ham}
\end{eqnarray}
where $c_i$ is a fermion annihilation operator at site $i$, and $X_{i}=\pm 1$ takes discrete values on the bond between sites $i$ and $i+1$. $t=1$ and $t'$ are the nearest-neighbor and next-nearest-neighbor hopping amplitudes of the fermions, respectively, $J$ is an Ising interaction between the nearest-neighbor $X_i$, and $U$ describes a local coupling between the fermions and $X_i$; see Fig. \ref{Fig:demonstration} for illustration. 

The thermal equilibrium properties of the system at temperature $\beta=1/k_B T$ are described by the density matrix $\hat\rho \propto e^{-\beta \hat{H}}$.
For a specific $X$ configuration, the Hamiltonian $\hat{H}(X)$ is non-Hermitian, yet the overall system remains Hermitian as we sum over $X$ configurations: 
\begin{equation}
    \hat{\rho} = \frac{1}{Z}\sum_X \hat{\rho}(X) = \frac{1}{Z}\sum_{-X} \hat{\rho}(-X) = \frac{1}{Z}\sum_X \hat{\rho}^\dagger(X) =  \hat{\rho}^\dagger,
    \label{eq:rhoherm}
\end{equation}
where we have used $\hat{H}(-X) = \hat{H}^\dagger(X)$ and thus $\hat{\rho}(-X) = \hat{\rho}^\dagger(X)$ for the fermions given the $X$ configuration. This derivation implicitly relies on the system's symmetry $D\rho D^{-1} = \rho$ under the configuration transformation $D:X\to -X$, which will spontaneously break at the phase transition.

 The partition function takes the form:
\begin{equation}
    Z = \sum_{X} \mbox{Tr}_c [e^{-\beta \hat{H}(X)}]
    = \sum_{X}  \prod_n [{1+e^{-\beta \varepsilon_n(X)}}],
    \label{eq:partitionf}
\end{equation}
where $\varepsilon_n(X)$  denotes the eigenvalues of $\hat{H}(X)$, which are either purely real or form complex-conjugate pairs due to the $\mathcal{P}\mathcal{T}$ symmetry of $\hat{H}(X)$\cite{supp}. For physical quantities' expectation values, we average over $X$ with the probability: 
\begin{equation}
    p({X}) = \frac{1}{Z}  \prod_n [{1+e^{-\beta \varepsilon_n(X)}}],
\end{equation}
 which remains positive-definite (due to the complex-conjugate-pair spectrum) and circumvents the MC sign problem.
Further, through a similarity transformation, $\hat{H}(X)$ for $t'=0$ maps to a translation-invariant Hatano-Nelson (HN) model, which depends only on the number $n_{\pm}$ of $X_i=\pm 1$ bonds; consequently, the fermion spectrum and weights are highly degenerate and the summation in Eq. \ref{eq:partitionf} can be carried out numerically and exactly \cite{supp}. Alternatively, we can analyze the model qualitatively via the self-consistently mean-field theory: 
\begin{eqnarray}
    \hat{H}_c &=& \sum_k \left[2t\cos(k)+2t'\cos(2k)+2iU\langle X\rangle \sin(k)\right] c^\dagger_k c_k, \nonumber \\ 
    \hat{H}_X &=& X \left[\frac{2iU}{L}\sum_k \langle c^\dagger_k c_k \rangle \sin(k) -2J\langle X\rangle \right],
    \label{eq:mft}
\end{eqnarray}
where we have neglected the fluctuations of $X$. For further details of methods, see the Supplemental Material \cite{supp}.

\begin{figure}
    \centering  
    \includegraphics[width=0.5\textwidth]{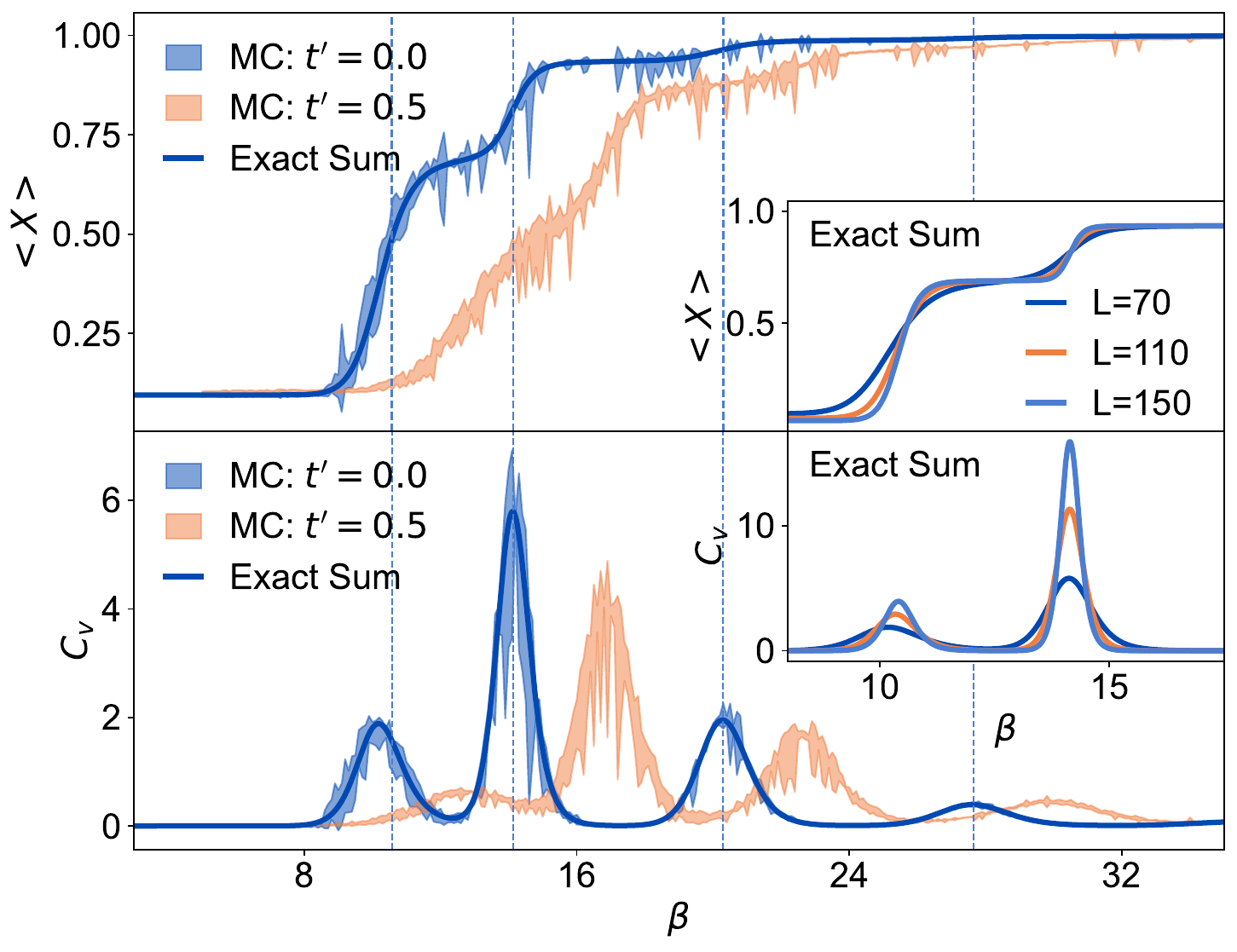}  
    \caption{The order parameter $\langle X\rangle$ and specific heat $C_V$ indicate the presence of SSB and long-range order in the 1D models in Eq. \ref{eq:ham} below a finite critical temperature $\beta>\beta_C$. The vertical dashed lines labels $\beta_C$ we pinpoint for $t'=0.0$ through finite-size analysis (inset). Here, $t=1$, $U=0.4$, $J=0.0$, $t'=0.0, 0.5$, and $L=70$. We provide consistent MC and exact results for $t'=0$, whereas only MC results are available at $t'=0.5$. }
    \label{Fig:classicalorderparameter}
\end{figure}

\begin{figure}
    \centering  
    \includegraphics[width=0.5\textwidth]{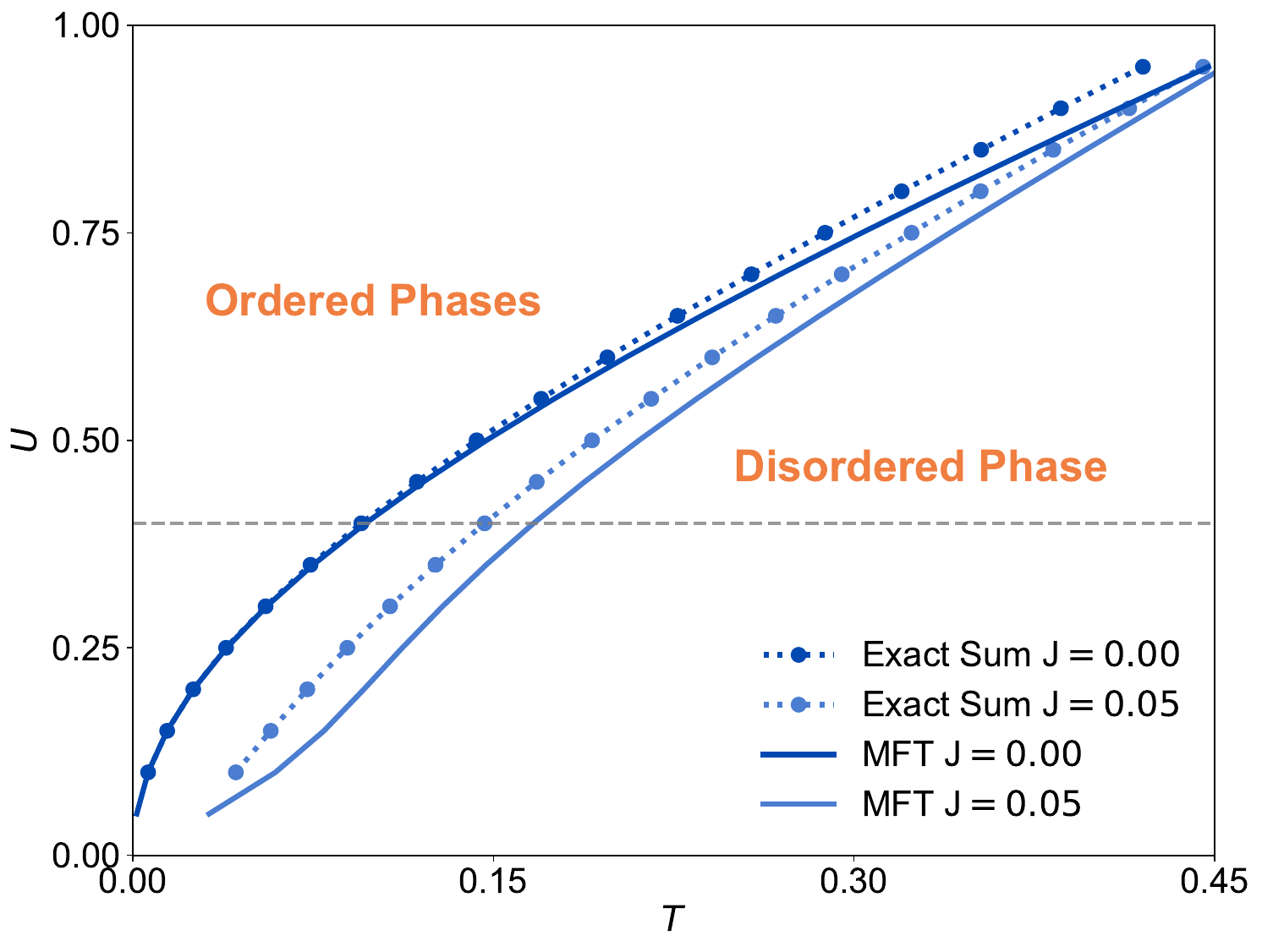}  
    \caption{The phase diagrams in $U$ and $T$ contain finite regions of ordered phases at low $T$ and large $U$. Here, $t'=0.0$ and $J=0, 0.05$. The horizontal gray dashed line follows the model parameters in Fig. \ref{Fig:classicalorderparameter}, which we use to determine $T_C$ ($\beta_C$). The solid curves are from the mean-field theory in Eq. \ref{eq:mft} and Ref. \cite{supp}. }
    \label{Fig:Phasediagram}
\end{figure}

\emph{Long-range order and energetics}--- We numerically evaluate physical properties such as the expectation value $\langle X\rangle=\langle |\sum_i X_i|\rangle/L$ and the specific heat $C_V$ over various system sizes and observe clear signatures of ordered phases at sufficiently low temperatures ($\beta>\beta_C$); see examples in Fig. \ref{Fig:classicalorderparameter}. Here, $\langle X \rangle$ serves as an Ising-type order parameter  and its long-range correlation emerges in the ordered phases \cite{supp}. 
The finite-size scaling of the specific heat $C_V$ accurately locates the transitions. These transitions exhibit first-order characteristics, as evident from the abrupt bifurcation of the lowest-free-energy configuration from zero to two opposite values in $X$ \cite{supp}, accompanied by spontaneous breaking of the $D$ symmetry through the selection of either of the two states with positive or negative $X$. We summarize the results in the phase diagrams in Fig. \ref{Fig:Phasediagram}, where the ordered phases hold finite regions leaning towards larger $U$ and lower $T$, indicating their sufficient robustness towards variations of parameters. 
 Beyond the primary transition, the system exhibits a sequence of phase transitions with distinctive signatures in both $\langle X\rangle$ and $C_V$, whose physical mechanism will be explained afterward. 
On the contrary, the entire parameter spaces consist of only a disordered phase under open boundary conditions (OBCs) or without non-Hermitian construction, e.g., $U=0$ or $U=0.4i$ \cite{supp}.

Such outcomes are counter-intuitive, as SSB is generally absent in 1D with local interactions and thermalization, and the disordered phase should prevail at any finite temperature. The underlying reason is that the energy cost $E$ of an isolated domain wall is approximately constant \footnote{The constant energy cost results from the discrete symmetry, such as $X_i=\pm 1$ here, while the continuous symmetry introduces gapless Goldstone modes, making the long-range order even more prone towards destruction. } and no match for its entropy $S$, which diverges with the system size $L\rightarrow\infty$ in the thermodynamic limit, in dominating the free energy $F=E-TS$ at arbitrarily low temperature $T$ \cite{Altland_Simons_2010}. Therefore, the domain walls proliferate, overthrow the long-range order, and forbid an SSB. Unlike previous 1D-SSB examples \cite{SSB1D1995, SSB1D1986, SSB1D1997, SSB1D1998} that circumvent the restrictions by staying far away from equilibrium, we make 1D SSB feasible via non-Hermitian construction's unique non-local and energetic properties. 

\begin{figure}
    \centering  
    \includegraphics[width=0.5\textwidth]{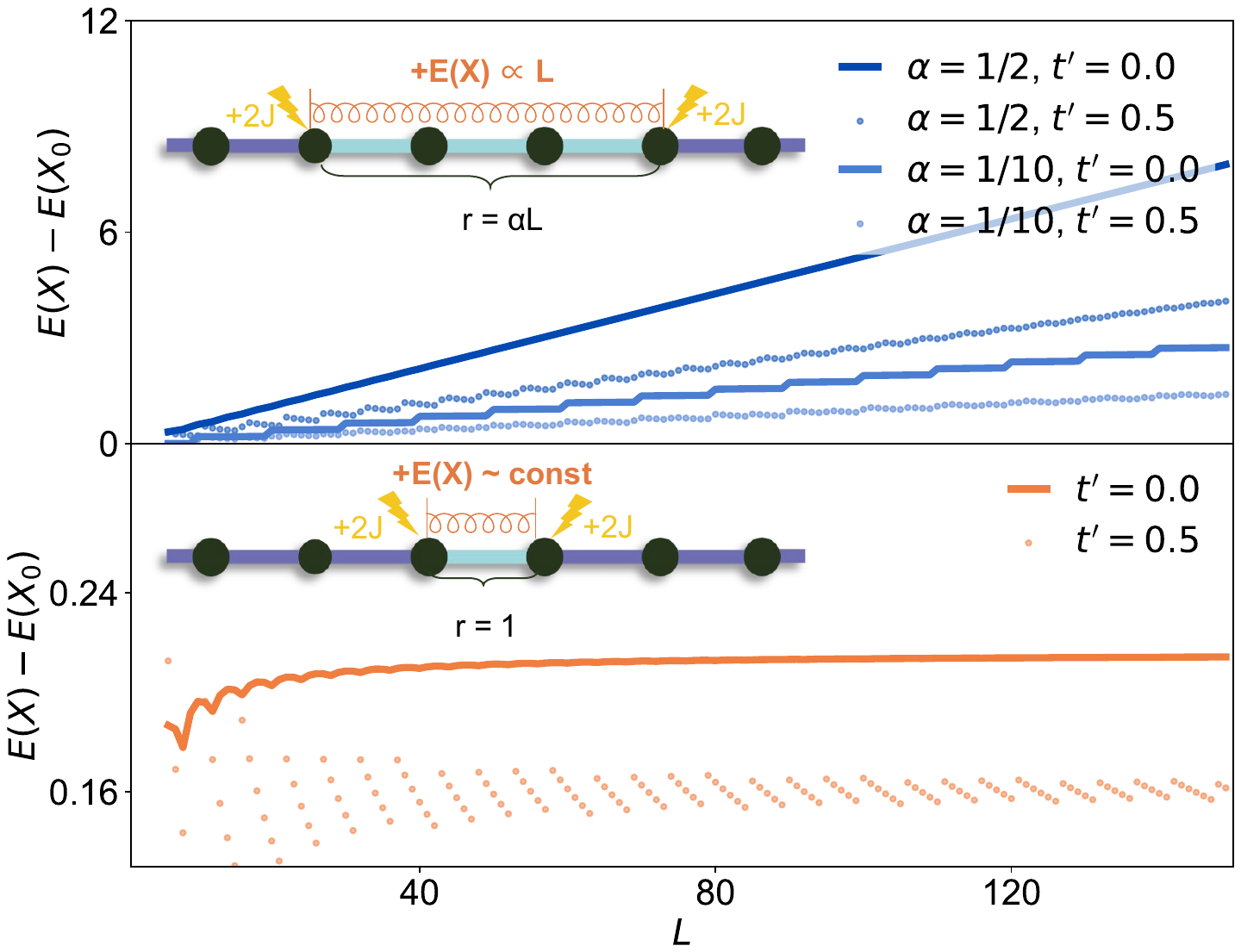}  
    \caption{The ground-state energy $E(X)$ increases linearly with two domain walls $r=\alpha L$ apart and saturates to a constant with nearby domain walls in the thermodynamic limit. Note the differences between the vertical scales. Here, $t'=0, 0.5$, $U=0.4$, and $X_0$ is uniformly ordered without domain walls.  The insets display schematic diagrams of the corresponding $X$ configurations featuring two domain walls, whose energy is attributed to both the fermions and the Ising interaction (a constant $2J$ energy cost) upon $X$.} 
    \label{Fig:domainwall}
\end{figure}

Let us consider the $t'=0$ case in Eq. \ref{eq:ham} as an example: given a specific $X$ configuration, the ground-state energy $E(X)$ of $\hat{H}(X)$ depends only on $n_{\pm}$; the minimum $E(X)$ corresponds to the uniform configuration $X_0$ with all $X_{i}=1$ ($X_{i}=-1$). Importantly, unlike the Ising interaction that contributes a constant energy cost of $2J$ for each domain wall, $E(X)$, through non-Hermitian construction, is no longer locally dependent on the (number of) domain walls in $X$, but on the separations between them as well. For instance, $E(X)$ increases linearly with two domain walls $r=\alpha L$ apart, while $E(X)$ saturates to a constant with nearby domain walls in the thermodynamic limit, as we illustrate in Fig. \ref{Fig:domainwall}. The trend remains at finite temperature. Such non-local energetic dependence applies to general scenarios beyond similarity transformations, e.g., $t' \neq 0$, but are absent in similar yet Hermitian constructions or under OBCs \cite{supp}.

Such emergent non-local behaviors are unique to non-Hermitian physics and differ qualitatively from conventional Hermitian constructions, whose energy cost is only generated at or around the boundaries or domain walls \footnote{The NHSE is an example of how boundaries and domain walls impose global impacts; the non-Hermitian delocalization also introduces non-locality under and only under PBCs. }. They are the key to the diminishing deterrence of SSB in 1D systems via non-Hermitian constructions: the domain walls are no longer free but strung together attractively proportional to their separations, prohibiting their global proliferation; despite recurrent fluctuations of domain-wall pairs in thermal equilibrium, they introduce merely local patches of opposing orders and pose no threat to the long-range order. We note that non-Hermitian constructions bear much potential and generalizability in modeling correlated physics beyond quasiparticles.

\begin{figure}
    \centering  
    \includegraphics[width=0.5\textwidth]{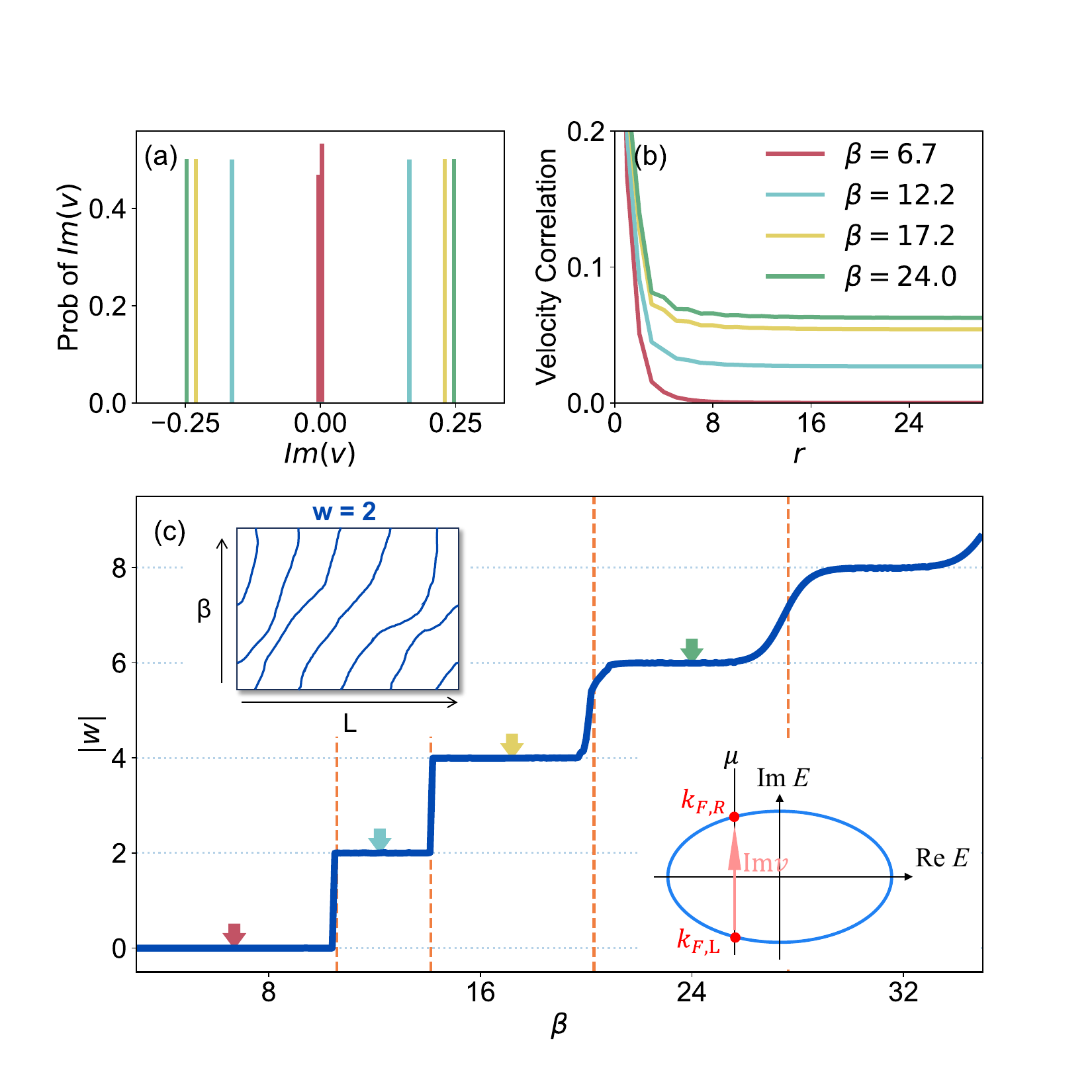}  
    \caption{The phases and phase transitions exhibit a topological nature from the fermion perspective.  (a): The probability distribution of $v$ shifts from peaking at 0 to finite-imaginary-value concentrations. (b): Long-range correlation of local $v$ emerges in the ordered phases. (c): The fermion path integral's winding number $w$ transitions from the conventional $|w|=0$ plateau (disordered phase) to quantized plateaus $|w|=2, 4, \cdots$ (ordered phase). The vertical dashed line labels the corresponding transitions in Fig. \ref{Fig:classicalorderparameter}; temperatures in (a-b) align with colored arrows in (c). Here, $U=0.4$, $J=0.0$, and $t'=0.0$. Inset in (c): (Upper right) an illustration of nontrivial winding number $w=2$ in (1+1)D path integral under PBCs, and (Lower left) an illustration of the imaginary velocity as from the complex energy spectrum, detailed in \cite{supp, hu2024residue}. }
    \label{Fig:Fermionproperties}
\end{figure}

\emph{Topological aspects of fermions}--- Here, we examine the SSB and ordered phases from a fermionic perspective. The semiclassical theory has suggested that the complex energy spectrum in non-Hermitian systems may induce a residual imaginary velocity \cite{hu2023nontrivial, hu2024residue}. In Fig. \ref{Fig:Fermionproperties}, we numerically compute the statistical distribution of such imaginary velocities, which exhibits a behavior analogous to the Ising-model magnetization across its SSB: (i) a peak at zero in the disordered phase and (ii) spontaneous bifurcation into conjugate imaginary sectors ($\pm \operatorname{Im} v$) in the ordered phases. Moreover, long-range correlations of local $v$ emerge in the ordered phases. While the global Hermiticity requires $\langle v \rangle = 0$, $\langle|v|\rangle$ emerges as a natural order parameter.

In quantum Monte Carlo simulations depicting imaginary time evolution within the (1+1)D path integral framework, the worldline winding number $w$ counts the number of times that the particle trajectories (worldlines) cross the periodic boundaries during an imaginary time $\beta=1/k_BT$. Conventionally, in Hermitian systems, the most relevant worldlines reside predominantly in the $w=0$ sector \cite{Sandvik1998winding}; however, this no longer holds for non-Hermitian systems under PBCs: the dominant worldlines winding number locates in $w \neq 0$ sector and is related to the imaginary velocity through $w=\beta \operatorname{Im}(\langle v\rangle)$ \footnote{This equation is taken from a previous result \cite{hu2024residue}, with the parameter $L$ absorbed into our definition of $\langle v \rangle$; see the Supplemental Material \cite{supp} for details. }, which reflects that the quantum dynamics of imaginary velocity induce unidirectional collective displacements in imaginary time \cite{supp, hu2023nontrivial, hu2024residue}. Here, we note that the SSB is also signaled and accompanied by the shift of dominance from the $w=0$ sector to the $w \neq 0$ sectors: while the disordered phase corresponds to a $|w|=0$ plateau, multiple order phases agree with respective nontrivial integer plateaus of winding numbers $|w|=2, 4, \cdots$, separated by abrupt changes at critical temperatures consistent with the phase transitions from $\langle X \rangle$ and $C_V$ in Fig. \ref{Fig:classicalorderparameter}. The even quantization in $|w|$ stems from the independent and simultaneous left and right Fermi-point contributions, and can be modified by de-synchronizing the Fermi-point contributions - for example, adding time-reversal symmetry-breaking terms to Eq. \ref{eq:ham} yields $\Delta |w|= \pm 1$ plateaus, see the Supplemental Material for details \cite{supp}. 

As the $D$ symmetry is intimately related to Hermiticity, the system exhibits non-Hermitian characteristics  after SSB, including finite imaginary velocity and non-zero winding numbers. It also leaves nontrivial marks on the fermion spectral functions: while singularities mainly concentrate on the real axis on the complex energy plane in the disordered phase, they materialize in imaginary regions on the complex energy plane in the order phases \cite{supp}. While spontaneous ``Hermiticity breaking" might appear unphysical, it bears a profound analogy to ``gauge-symmetry breaking" in superconductors - the ``breaking" is only explicit at a mean-field level, as exemplified by a finite $\langle X\rangle$ in Eq. \ref{eq:mft} \cite{supp}.

\emph{Conclusions and discussion}--- Previous studies have found that Hermitian physics may emerge from non-Hermitian models or become influenced by non-Hermitian fixed points \cite{zixiang2024}. Here, we have shown that non-Hermitian construction may also impact the resulting Hermitian systems nontrivially, validating SSB even in one dimension, local interaction, and thermal equilibrium. On the one hand, such anomalies are due to the implicit non-locality induced by non-Hermitian physics, which correlates domain walls and upholds the long-range order; on the other hand, the phases and phase transitions are essentially topological from a quantum perspective, characterized by global integers of the fermion path-integral windings. Our numerical calculations have also provided accuracy and consistent results on the nature of such phases and SSB transitions. Thus, such non-Hermitian constructions carry novel prospects for SSB and non-Hermitian physics. 

Non-Hermitian constructions are generalizable beyond our model examples to scenarios consisting of fermion or boson fields instead of $X$. The coupling similar to non-Hermitian $\hat{H}(X)$ will help introduce nontrivial, non-local quasiparticle energetics and thus establish exotic effective theory beyond free quasiparticles. Also, we obtain interacting, purely fermionic models by integrating out the $X_i=\pm 1$ degrees of freedom, which category we will discuss further in a separate future work. Interestingly, we have also demonstrated lively examples of a novel type of boundary conditions with a specific winding number $w$, where the $w=0$ case is simply OBCs \footnote{Such boundary conditions are dual to the flux phases $e^{i\phi}$ across PBCs. }. Such constructions and boundary conditions offer a unique and non-perturbative variational dimension, which may significantly impact the universal physics of condensed matter systems \cite{supp}.

\emph{Acknowledgment:} We acknowledge insightful discussions with Yongxu Fu and Shi-Xin Hu, and generous support from the National Natural Science Foundation of China (Grants No.92270102 \& No.12174008) and the National Key R\&D Program of China (Grant No.2022YFA1403700).

\bibliography{ref.bib}

\end{document}